\documentclass[a4paper,10pt]{article}
\begin{document}

\makeatletter \@addtoreset{equation}{section} \makeatother
\renewcommand{\theequation}{\thesection.\arabic{equation}}
\baselineskip 15pt 

\title{\bf A general scheme for ensemble purification}

\author{Angelo Bassi\footnote{e-mail: bassi@ictp.trieste.it}\\
{\small The Abdus Salam International Centre for Theoretical
Physics, Trieste, Italy} \\ {\small and Istituto Nazionale
di Fisica Nucleare, sezione di Trieste, Italy}\\ and \\
\\ GianCarlo Ghirardi\footnote{e-mail: ghirardi@ts.infn.it}\\
{\small Department of Theoretical Physics of the University of
Trieste,}\\ {\small the Abdus Salam International Centre for
Theoretical Physics, Trieste, Italy,} \\ {\small and Istituto
Nazionale di Fisica Nucleare, sezione di Trieste, Italy}}
\date{}
\maketitle
\begin{abstract}
We exhibit a general procedure to purify any given ensemble by
identifying an appropriate interaction between the physical system
$S$ of the ensemble and the reference system $K$. We show that the
interaction can be chosen in such a way to lead to a spatial
separation of the pair $S$--$K$. As a consequence, one can use it
to prepare at a distance different equivalent ensembles. The
argument associates a physically precise procedure to the purely
formal and fictitious process usually considered in the
literature. We conclude with an illuminating example taken from
quantum computational theory.
\end{abstract}

\section{Introduction}

A statistical ensemble ${\cal E}$ of physical systems $S$ is
characterized by a (finite, countable or continuous) set of
positive numbers $p_{i}$ summing up to 1 and by a corresponding
set of normalized vectors $\vert \psi_{i}\rangle$ of the Hilbert
space ${\cal H}^{S}$ associated to the system $S$, so that  we
will write ${\cal E}(p_{i},\vert \psi_{i}\rangle)$ to represent
it. The statistical operator $\rho_{\cal E}$ (a trace--class,
trace one, semipositive definite operator) associated to ${\cal
E}(p_{i},\vert \psi_{i}\rangle)$ is defined as:
\begin{equation}  \label{eq1}
\rho_{{\mathcal E}}=\sum_{i}
p_{i}\vert\psi_{i}\rangle\langle\psi_{i}\vert.
\end{equation}

A point of great conceptual relevance which marks a radical
difference between  the classical and quantum cases is that, while
in classical mechanics the assignment of the statistical operator
$\rho({\bf r,p})$ uniquely identifies the ensemble, within quantum
mechanics, as it is well known, the correspondence between
statistical ensembles and statistical operators is infinitely many
to one.

With reference to this point, let us consider the set of all
statistical ensembles of systems like the one under consideration.
Such a set can be naturally endowed with an equivalence relation.
\begin{quote}
{\bf Definition}:{\it We will say that two statistical ensembles
${\cal E}$ and ${\cal E^{*}}$ are equivalent, and we will write
${\cal E}\equiv{\cal E^{*}}$, iff $\rho_{{\cal E}}=\rho_{\cal
E^{*}}$}.
\end{quote}
It is obvious that the just defined relation is reflexive,
symmetric and transitive and that it leads to a decomposition of
the set of all ensembles into disjoint equivalence classes. We
will denote as $[{\cal E}]$ the equivalence class containing the
ensemble ${\cal E}$.

Purification of an ensemble \cite{nielsen} is a procedure by which
one associates to the ensemble a pure state $\vert \Psi\rangle$ of
an appropriately enlarged Hilbert space ${\cal H}^{S+K}={\cal
H}^{S}\otimes{\cal H}^{K}$, where $K$ is a {\it reference} system
whose Hilbert space ${\cal H}^{K}$ we assume to be
infinite--dimensional for reasons which will become clear in a
moment. The fundamental request on $\vert \Psi\rangle$ is that, by
measuring an appropriate observable of $K$ and confining attention
to the system $S$ alone, one can prepare the desired ensemble
${\cal E}(p_{i},\vert \psi_{i}\rangle)$.

The first proof that, given two equivalent ensembles ${\cal
E}(d_{i},\vert \phi_{i}\rangle)$ and ${\cal E}(p_{j},\vert
\chi_{j}\rangle)$, one can find two orthonormal sets $\{\vert
A_{i}\rangle\}$ and $\{\vert B_{j}\rangle\}$ of ${\cal H}^{K}$
such that
\begin{equation}
\vert \Psi\rangle=
\sum_{i}\sqrt{d_{i}}\vert\phi_{i}\rangle\otimes\vert
A_{i}\rangle=\sum_{j} \sqrt{p_{j}}\vert\chi_{j}\rangle\otimes\vert
B_{j}\rangle
\end{equation}
has been exhibited by Gisin \cite{gisin}. This result is
particularly relevant since it is related to the request that no
faster--than--light signals can be send between distant observers.

Subsequently, Hughston {\it et al.} \cite{hughston} have
generalized the above result, providing a complete classification
of equivalent ensembles: using the purification procedure, they
have derived necessary and sufficient conditions for two ensembles
to be equivalent.

In the literature (see, e.g., \cite{nielsen}), ensemble
purification is usually considered as a purely mathematical tool:
one does not identify any dynamical mechanism which could be used
to actually implement it, and the system $K$ is considered a
fictitious system without a direct physical significance. The aim
of this paper is to exhibit a precise physical procedure in order
to purify any ensemble by making the system $S$ interact with a
system $K$, in such a way that the desired pure state
$|\Psi\rangle$ be actually produced. Then one can use it to
prepare any desired ensemble of the equivalence class.

\section{Statistical ensembles and the purification process:
the constructive procedure}

As remarked above, it is our purpose to present a formal
constructive mechanism to purify any given ensemble, showing at
the same time how, by resorting to this procedure, one can use the
obtained pure state to generate all ensembles of systems $S$
equivalent to the one one has purified. The procedure is based on
a formalism which parallels strictly the one proposed by von
Neumann for implementing ideal measurement processes of the first
kind, even though the system $K$, which plays a role analogous to
the one of the measuring apparatus in his treatment, can very well
be (and actually we will consider it to be) a microsystem.

Our starting point is the consideration of an equivalence class
$[{\cal E}]$ of ensembles of systems $S$. Within such a class
there is the ensemble ${\cal E}(d_{i},\vert \phi_{i}\rangle)$
which corresponds to the spectral decomposition of the associated
statistical operator having the positive numbers $d_{i}$ as
eigenvalues and the $\vert \phi_{i}\rangle$ as the associated
orthonormal eigenvectors. Such a decomposition is unique, apart
from accidental degeneracies which, if they occur, can be disposed
of as one wants, so that we will consider the eigenvectors
$|\phi_{i}\rangle$ as precisely assigned vectors. We assume that
the index $i$ runs from $0$ to $n$, without committing ourselves
about the fact that $n$ is finite or infinite and about the fact
that the orthonormal set $\{\vert\phi_{i}\rangle\}$ be a complete
set of ${\cal H}^{S}$ or not.

Let us consider now the orthonormal states $\vert\phi_{i}\rangle$
and let us assume that there exist a physical system $K$, whose
associated Hilbert space ${\cal H}^{K}$  is infinite dimensional,
a state $\vert a_{0}\rangle$ of ${\cal H}^{K}$ and an interaction
hamiltonian $H^{S+K}$ of  ${\cal H}^{S+K}$ such that the $S$--$K$
interaction lasting for a certain time interval $T$ induces the
following evolution:
\begin{equation} \label{eq3}
\vert\phi_{i}\rangle\otimes\vert
a_{0}\rangle\Longrightarrow\vert\phi_{i}\rangle\otimes\vert
a_{i}\rangle,\;\;\;\langle a_{i}\vert a_{j}\rangle= \delta_{ij},
\end{equation}
where $|a_{i}\rangle$ are statevectors belonging to ${\mathcal
H}^{K}$.

In the next section we will exhibit a simple hamiltonian having
such a property and leading also to an arbitrarily chosen
separation in space of the systems $S$ and $K$. We stress that we
need ${\cal H}^{K}$ to be infinite dimensional if we want to be
able to build a state $\vert\Psi_{T}\rangle$ which will allow us
to prepare any ensemble whatsoever in the equivalence class under
consideration by measurement procedures on systems $K$, since in
any equivalence class there are always ensembles containing an
infinite number of states.

Given the ensemble ${\cal E}(d_{i},\vert \phi_{i}\rangle)$ we
consider the state:
\begin{equation} \label{eq4}
\vert\Psi_{0}\rangle=\sum_{i=1}^n\sqrt{d_{i}}\vert
\phi_{i}\rangle\otimes\vert a_{0}\rangle,
\end{equation}
and we let it evolve through the interval $T$. According to Eq.
(\ref{eq3}) and due to the linearity of the quantum evolution, we
get
\begin{equation} \label{eq5}
\vert\Psi_{0}\rangle\Longrightarrow\vert\Psi_{T}\rangle=
\sum_{i=1}^n\sqrt{d_{i}}\vert
\phi_{i}\rangle\otimes\vert a_{i}\rangle.
\end{equation}
We consider now an arbitrary complete orthonormal set $\{\vert
b_{j}\rangle\}$, $j=0,1,...,\infty$, of ${\cal H}^{K}$ and we
complete (if necessary) the set $\{\vert a_{i}\rangle\}$,
$i=0,1,...,n$, to a set $\{\vert A_{i}\rangle\}$ by adding to it
orthonormal states spanning the manifold of ${\cal H}^{K}$
orthogonal to the one generated by the $\{\vert a_{i}\rangle\}$
themselves. Obviously we have:
\begin{equation}
\vert A_{i}\rangle=\sum_{j=0}^{\infty} U_{ij}\vert
b_{j}\rangle,\;\;\;i=0,1,...,\infty,
\end{equation}
where $U_{ij}$ is a unitary matrix of ${\cal H}^{K}$. From Eq.(5)
we get:
\begin{eqnarray} \label{eq7}
\vert\Psi_{T}\rangle &  = & \sum_{i=0}^n\sum_{j=0}^{\infty}
\sqrt{d_{i}}\vert \phi_{i}\rangle \otimes U_{ij}\vert b_{j}\rangle
\nonumber \\
& = & \sum_{j=0}^{\infty} (\sum_{i=0}^n \sqrt{d_{i}}U_{ij}\vert
\phi_{i}\rangle)\otimes\vert b_{j}\rangle \nonumber \\
& = & \sum_{j=0}^{\infty}\vert\tilde{\chi_{j}}\rangle\otimes\vert
b_{j}\rangle.
\end{eqnarray}
Note that $\langle\Psi_{T}\vert\Psi_{T}\rangle=1$ implies:
\begin{equation}
\sum_{j,k=0}^{\infty}\langle\tilde{\chi_{j}}\vert\langle
b_{j}\vert
b_{k}\rangle\vert\tilde{\chi_{k}}\rangle=
\sum_{j=0}^{\infty}\|\vert\tilde{\chi
_{j}}\rangle\|^2=1.
\end{equation}
The states $\vert\tilde{\chi_{j}}\rangle$ are not normalized, so
that, putting
$\vert\chi_{j}\rangle=
\vert\tilde{\chi_{j}}\rangle/\|\vert\tilde{\chi_{j}}\rangle\|$,
we have:
\begin{equation} \label{eq9}
\vert\Psi_{T}\rangle=
\sum_{j=0}^{\infty}\|\vert\tilde{\chi_{j}}\rangle\|\vert{\chi_{j}}
\rangle\otimes\vert b_{j}\rangle.
\end{equation}

If we measure now an observable of the system $K$ having a
non--degenerate spectrum with $\vert b_{j}\rangle$ as eigenvectors
and we confine our attention to the resulting ensemble of systems
$S$, we obtain the ensemble ${\cal
E}(\|\vert\tilde{\chi_{j}}\rangle\|^{2},\vert \chi_{j}\rangle)$.
Note that since both $\sum_{i=0}^n
d_{i}\vert\phi_{i}\rangle\langle\phi_{i}\vert$ and
$\sum_{j=0}^{\infty}\|\vert\tilde{\chi_{j}}\rangle\|^{2}
\vert\chi_{j}\rangle\langle\chi_{j}\vert $ are obtained by taking
the partial trace on ${\cal H}^{K}$ of
$\vert\Psi_{T}\rangle\langle\Psi_{T}\vert$, they are equal and the
corresponding ensembles belong to the same equivalence class. Thus
we have proved that starting from the state (\ref{eq5}) and
choosing an observable having $|b_{j}\rangle$ as eigenstates, we
generate an ensemble which belongs to the same equivalence class
of the original one.

The relevant question we have to face now is the following: can
all statistical ensembles belonging to the equivalence class of
${\cal E}(d_{i},\vert \phi_{i}\rangle)$ be obtained by properly
choosing the observables of the system K we are going to measure?
The answer is yes, as it is easily proved. To this purpose, let us
consider an arbitrary ensemble ${\cal E}(p_{j},\vert
\tau_{j}\rangle)$ equivalent to ${\cal E}(d_{i},\vert
\phi_{i}\rangle)$; we suppose that the index $j$ runs from $0$ to
$N$ ($\geq n$), without excluding the case in which $N$ is
infinite. We know that the fact that the statistical operators
associated to such ensembles are identical implies that the
normalized states $\vert\tau_{j}\rangle$ are linear combinations
of the orthonormal states $\vert\phi_{i}\rangle$:
\begin{equation} \label{eq10}
\vert\tau_{j}\rangle=\sum_{i=0}^n
b_{ji}\vert\phi_{i}\rangle,\;\;\;j=0,1,...,N.
\end{equation}
We define now a rectangular matrix $V_{ij}$ having $n+1$ rows and
$N+1$ columns by putting:
\begin{equation} \label{defv}
V_{ij}=\sqrt{\frac{p_{j}}{d_{i}}}\,b_{ji},
\;\;\;i=0,1,...,n;\;\;j=0,1,...,N.
\end{equation}
From the relation $\sum_{i=0}^n
d_{i}|\phi_{i}\rangle\langle\phi_{i}|=\sum_{j=0}^{N}
p_{j}|\tau_{j}\rangle\langle\tau_{j}|$, using Eq. (10) we
immediately get:
\begin{equation}
\sum_{j=0}^{N} p_{j}b_{ji}b_{jk}^{*}=d_{i}\delta_{ik}.
\end{equation}
The above relation implies:
\begin{equation} \label{eq13}
\sum_{j=0}^{N}V_{ij}(V^{\dagger})_{jk}=
\sum_{j=0}^{N}\sqrt{\frac{p_{j}}{{d_{i
}}}}\,\sqrt{\frac{p_{j}}{d_{k}}}\,b_{ji}b_{jk}^{*}=\delta_{ik}.
\end{equation}
We thus have $n+1$ normalized and orthogonal vectors
$\{\tilde{{\mathbf w}}_{r}\}$, $r=0,1,...,n$ of ${\bf C}^{N+1}$,
whose components are the row elements of the matrix $V_{rj}$:
\begin{equation}
\tilde{{\bf w}}_{r}=(V_{r0},V_{r1},...V_{rN}),\;\;\;r=0,1,...,n.
\end{equation}
If $N$ is finite, we pass from the vectors $\{ \tilde{\bf w}_{r}
\}$ to new vectors $\{ {\bf w}_{r} \}$ of ${\bf C}^{\infty}$ by
considering equal to zero the components of $\{ {\bf w}_{r} \}$
from $N+1$ on. We then extend the set $\{{\bf w}_{r}\}$ to a
complete orthonormal set of ${\bf C}^{\infty}$, by adding
appropriately chosen normalized vectors $\{{\bf w}_{s}\}$,
$s=n+1,..,\infty$. Correspondingly, the rectangular matrix
$V_{ij}$ of Eq. (\ref{defv}) is transformed into an infinite
square matrix, whose rows are the components of the vectors
$\{{\bf w}_{r}\}$, for $r=0,1,..,\infty$. Due to Eq. (\ref{eq13})
and the procedure we have followed, this infinite square matrix
--- which we keep calling $V_{ij}$ --- is unitary.

Let us consider now an observable $\Omega^{K}$ of ${\mathcal
H}^{K}$ having a purely discrete and non degenerate spectrum with
eigenvectors $|B_{j}\rangle = V^{\dagger}_{ji}|A_{i}\rangle$; this
implies that $|A_{i}\rangle = V_{ij}|B_{j}\rangle$. Since $V_{ij}$
is unitary, we can repeat the previous procedure which amounts
simply in replacing, in Eq. (\ref{eq5}), the states $|a_{i}\rangle
= |A_{i}\rangle$ appearing there with their Fourier expansion in
terms of the set $\{ |B_{j}\rangle \}$. Then Eq. (\ref{eq5}) takes
the form (\ref{eq9}) where, according to the definition of
$|\tilde{\chi}_{j}\rangle$ given in Eq. (\ref{eq7}) and of
$|\tau_{i}\rangle$ given in Eq. (\ref{eq10}):
\begin{eqnarray}
|\tilde{\chi}_{j}\rangle & = & \sum_{i=0}^{n} \sqrt{d_{i}}\,
V_{ij}\, |\phi_{i}\rangle = \sqrt{p_{j}} \sum_{i=0}^{n}\, b_{ji}\,
|\phi_{i}\rangle = \nonumber \\
& = & \sqrt{p_{j}}\, |\tau_{j}\rangle.
\end{eqnarray}

This shows that $\| |\tilde{\chi}_{j}\rangle \|^{2} = p_{j}$ and
that normalizing $|\tilde{\chi}_{j}\rangle$ we get the states
$|\tau_{j}\rangle$. Accordingly, Eq. (\ref{eq9}) becomes:
\begin{equation}
|\Psi_{T}\rangle = \sum_{j=0}^{N} \sqrt{p_{j}}\,
|\tau_{j}\rangle\, \otimes \, |B_{j}\rangle,
\end{equation}
so that measurement of $\Omega^{K}$ reduces the state
$|\Psi_{T}\rangle$ to the desired ensemble ${\cal E}(p_{j},
|\tau_{j}\rangle)$. Since ${\mathcal E}(p_{j}, |\tau_{j}\rangle)$
is an arbitrary ensemble belonging to the equivalence class
$[{\mathcal E}(d_{i}, |\phi_{i}\rangle)]$, this completes our
proof.

The now obtained result shows how, once one has prepared the pure
state $|\Psi_{T}\rangle$, he has an immediate complete
classification of all ensembles belonging to the equivalence class
of ${\cal E}(d_{i}, |\phi_{i}\rangle)$, an alternative way of
deriving the nice result of \cite{hughston}.

Of course, within any equivalence class there are also mixtures
which involve a continuous union of pure states i.e.
\begin{equation}
{\mathcal E}(p(\lambda), |\phi_{\lambda}\rangle) \;
\longrightarrow \; \rho_{\mathcal E} = \int d\lambda\,
p(\lambda)\, |\phi_{\lambda}\rangle\langle\phi_{\lambda}|,
\end{equation}
with $\int d\lambda\, p(\lambda) = 1$. To get such mixtures from
the pure state $|\Psi\rangle$ we have, obviously, to measure with
infinite precision an observable of ${\mathcal H}^{K}$ having a
continuous spectrum. This is formally but not practically
feasible.

Concluding, if we can implement our ``von Neumann--like ideal
interaction scheme" we can perform the desired purification and
then prepare any one of the ensembles in the equivalence class of
${\cal E}(d_{i},| \phi_{i}\rangle)$ by performing an appropriate
measurement on the system $K$.

\section{The appropriate hamiltonian for the desired purification}

To face our problem let us consider the following self-adjoint
operator of ${\cal H}^{S+K}$:
\begin{equation}
H_{j}=i\,| \phi_{j}\rangle\langle \phi_{j}|\otimes[|
a_{0}\rangle\langle a_{j}|-| a_{j}\rangle\langle a_{0}|],
\end{equation}
and let us evaluate its powers. We have:
\begin{eqnarray}
H_{j}^{2n+1} & = & H_{j}, \nonumber \\
H_{j}^{2n} & = & | \phi_{j}\rangle\langle \phi_{j}|\otimes[|
a_{0}\rangle\langle a_{0}|+| a_{j}\rangle\langle a_{j}|].
\end{eqnarray}

Let us consider now the operator $\exp(-i\omega H_{j}T)$:
\begin{equation}
\exp(-i\omega H_{j}T)=\cos(\omega H_{j}T)-i\sin(\omega H_{j}T).
\end{equation}
Since $\sin$ contains only odd powers of $H_{j}$ we have:
\begin{equation}
\sin(\omega H_{j}T)=H_{j}\sin(\omega T),
\end{equation}
while, since all even powers of $H_{j}$ equal $H_{j}^{2}$ we can
write:
\begin{eqnarray}
\cos(\omega H_{j}T) & = &
1-H_{j}^{2}[-1+1+\frac{1}{2}\omega^{2}T^{2}-...] \nonumber \\
& = & 1-H_{j}^{2}+H_{j}^{2}\cos(\omega T).
\end{eqnarray}

We now choose for $T$ a value such that $\cos(\omega
T)=0,\;\;\sin(\omega T)=1$, getting:
\begin{equation}
\exp(-i\omega H_{j}T)=1-H_{j}^{2}-iH_{j}.
\end{equation}
The last equation implies that
\begin{eqnarray}
\exp(-i\omega H_{j}T)\,|\phi_{j}\rangle\otimes| a_{0}\rangle & = &
[1-H_{j}^{2}-iH_{j}]\,|\phi_{j}\rangle \otimes | a_{0}\rangle
\nonumber \\ & = &|\phi_{j}\rangle\otimes| a_{j}\rangle,
\end{eqnarray}
as desired.

We remark now that $[H_{j},H_{k}]=0$ and
$H_{k}|\phi_{j}\rangle\otimes| a_{0}\rangle=0$ for $k\neq j$.
Accordingly, if consideration is given to the hamiltonian
$H=\sum_{j=0}^{\infty}H_{j}$ we have:
\begin{equation}
\exp(-i\omega HT)\,|\phi_{j}\rangle\otimes|
a_{0}\rangle=|\phi_{j}\rangle\otimes| a_{j}\rangle,\;\;\;
\forall{j}.
\end{equation}
Therefore, we have explicitly exhibited an hamiltonian which
performs our game, i.e., it leads to the desired purification of
our statistical mixture.

Actually, the purification procedure becomes interesting when one
can prepare a desired mixture among all those of an equivalence
class at--a--distance, as appropriately stressed by Gisin
\cite{gisin}. To reach this goal a very small change in our
formalism is necessary. Let us identify the states $|
a_{j}\rangle$ of our equation with the internal eigenstates of a
system (e.g. the stationary states of an hydrogen atom). One can
then add to our hamiltonian a term $\gamma P_{CM}$, where $\gamma$
is an appropriately chosen c-number and $P_{CM}$ is the
center--of--mass momentum of the system. The evolution induced by
the total hamiltonian implies a displacement of the system $K$ of
an amount governed by the value of $\gamma$, so that in the time
interval $T$ it is brought arbitrarily far from the space region
where its interaction with $S$ took place. In brief, the auxiliary
system is far apart and one can actually use the pure state to
prepare the desired statistical ensemble of systems $S$
at--a--distance.

\section{A quantum computational example}

It is interesting to notice that, for most cases of interest in
quantum computational theory, the outlined procedure can be easily
implemented by resorting to elementary logical gates. To this
purpose, let us suppose that $S$ is a qubit, i.e. a two--level
system, and let us denote as $|0\rangle$, $|1\rangle$ the
computational basis of ${\mathcal H}^{S}$. The controlled--{\small
NOT} operator acting on qubit $S$ (taken as the control bit) and
on the two--dimensional manifold spanned by the computational
basis states $|a_{0}\rangle$, $|a_{1}\rangle$ of system $K$ (taken
as the target bit) induces precisely the transformation:
\begin{eqnarray}
|0\rangle \otimes |a_{0}\rangle & \Longrightarrow & |0\rangle
\otimes |a_{0}\rangle \nonumber \\
|1\rangle \otimes |a_{0}\rangle & \Longrightarrow & |1\rangle
\otimes |a_{1}\rangle,
\end{eqnarray}
which is the desired evolution. In this way, we can purify any
statistical ensemble belonging to the equivalence class of
\begin{equation}
{\cal E}(p, |0\rangle; 1-p, |1 \rangle), \qquad 0 < p < 1,
\end{equation}
by starting with an appropriate superposition analogous to the one
of Eq. (\ref{eq4}).

Let us now consider an arbitrary equivalence class, different from
the previous one and containing the ensemble (corresponding to the
diagonal form of $\rho$):
\begin{equation}
{\cal E}(q, |x_{+}\rangle; 1-q, |x_{-}\rangle), \qquad 0 < q < 1,
\end{equation}
where $|x_{+}\rangle$ and $|x_{-}\rangle$ are a basis obtained
from the computational basis $|0\rangle$, $|1\rangle$ by an
appropriate ``rotation'' of the system:
\begin{equation}
|x_{+}\rangle \; = \; R_{S}\, |0\rangle, \qquad |x_{-}\rangle \; =
\; R_{S}\, |1\rangle.
\end{equation}
The circuit that implements the evolution
\begin{eqnarray}
|x_{+}\rangle\otimes |a_{0}\rangle & \Longrightarrow &
|x_{+}\rangle\otimes|a_{0}\rangle, \nonumber \\
|x_{-}\rangle\otimes |a_{0}\rangle & \Longrightarrow &
|x_{-}\rangle\otimes|a_{1}\rangle,
\end{eqnarray}
leading to the purification of the ensemble, corresponds to a
``rotation'' $R_{S}^\dagger$ on the control bit, followed by a
controlled--{\small NOT} gate and by an inverse ``rotation''
$R_{S}$, as shown in the picture.
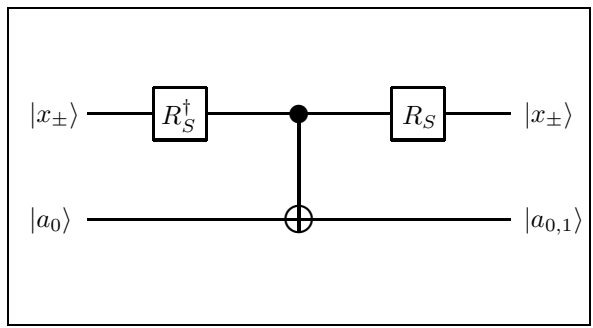
\begin{center}
\begin{picture}(220,120)(0,0)
\put(0,0){\line(1,0){220}} \put(0,120){\line(1,0){220}}
\put(0,0){\line(0,1){120}} \put(220,0){\line(0,1){120}}
\thicklines \put(30,80){\line(1,0){25}}
\put(75,80){\line(1,0){70}} \put(165,80){\line(1,0){25}}
\put(30,40){\line(1,0){160}}
\put(55,70){\line(1,0){20}} \put(55,90){\line(1,0){20}}
\put(55,70){\line(0,1){20}} \put(75,70){\line(0,1){20}}
\put(145,70){\line(1,0){20}} \put(145,90){\line(1,0){20}}
\put(145,70){\line(0,1){20}} \put(165,70){\line(0,1){20}}
\put(110,80){\circle*{7}} \put(110,80){\line(0,-1){45}}
\put(110,40){\circle{10}}
\put(58,76){$R_{S}^\dagger$} \put(149,76){$R_{S}$}
\put(8,77){$|x_{\pm}\rangle$} \put(8,37){$|a_{0}\rangle$}
\put(195,77){$|x_{\pm}\rangle$} \put(195,37){$|a_{0,1}\rangle$}
\end{picture}
\end{center}
Thus, the Hamiltonian that induces the desired evolution can be
identified with a rotation in ${\mathcal H}^{S}$, a
controlled--{\small NOT} operation in ${\mathcal
H}^{S}\otimes{\mathcal H}^{K}$, and finally a counter--rotation in
${\mathcal H}^{S}$.

In this way, we have identified the appropriate way to purify any
statistical ensemble of the two--dimensional system $S$. Useless
to say, our procedure can be easily generalized to systems
containing several qubits and, more in general, to arbitrary
quantum systems.

\end{document}